\documentclass[aps]{revtex4}
\usepackage{epsfig}
\usepackage{amsmath,amssymb}

\parskip=\medskipamount

\newcommand{\eq}[1]{(\ref{#1})}
\newcommand{\fig}[1]{Fig.\ref{#1}}

\newcommand{\be}{\begin{equation}}
\newcommand{\ee}{\end{equation}}

\newcommand{\barr}{\begin{array}}
\newcommand{\earr}{\end{array}}

\newcommand{\beqn}{\begin{eqnarray}}
\newcommand{\eeqn}{\end{eqnarray}}

\newcommand{\bs}{\begin{subequations}}
\newcommand{\es}{\end{subequations}}

\newcommand{\bw}{\begin{widetext}}
\newcommand{\ew}{\end{widetext}}

\newcommand\disp{\displaystyle}

\newcommand{\la}{\left<}
\newcommand{\ra}{\right>}

\begin{document}

\title{Bethe Ansatz in the Bernoulli Matching Model of Random Sequence Alignment}

\author{Satya N. Majumdar$^1$, Kirone Mallick$^2$,  and Sergei Nechaev$^1$\footnote{Also at: P.N.
Lebedev Physical Institute of the Russian Academy of Sciences, 119991, Moscow, Russia}}

\affiliation{$^1$Laboratoire de Physique Th\'eorique et Mod\`eles
Statistiques,\\ Universit\'e de Paris-Sud, CNRS UMR 8626, 91405 Orsay Cedex, France \\
$^2$ Service de Physique Th\'eorique, Saclay, 91191 Gif-sur-Yvette cedex, France}

\date{\today}

\begin{abstract}

For the Bernoulli Matching model of sequence alignment problem we apply the Bethe ansatz technique
via an exact mapping to the 5--vertex model on a square lattice. Considering the terrace--like
representation of the sequence alignment problem, we reproduce by the Bethe ansatz the results for
the averaged length of the Longest Common Subsequence in Bernoulli approximation. In addition, we
compute the average number of nucleation centers of the terraces.

\noindent


\end{abstract}

\maketitle

\section{Introduction: Bernoulli Matching model of Sequence Alignment}
\label{sect:2}

The goal of a sequence alignment problem is to find similarities in patterns in different
sequences. Sequence alignment is one of the most useful quantitative methods of evolutionary
molecular biology \cite{W1,Gusfield,DEKM}. A classic alignment problem deals with the search of the
Longest Common Subsequence  (LCS) in two random sequences. Finding analytically the statistics of
LCS of a pair of sequences randomly drawn from the alphabet of $c$ letters is a challenging problem
in computational evolutionary biology. The exact asymptotic results for the distribution of LCS
have been derived recently in \cite{MN2} in a simpler, yet nontrivial, variant called the Bernoulli
Matching (BM) model (see details below). It has been shown in \cite{MN2} via a sequence of mappings
that in the BM model, for all $c$, the distribution of the asymptotic length of the LCS, suitably
scaled, is identical to the Tracy--Widom distribution of the largest eigenvalue of a random matrix
whose entries are drawn from a Gaussian Unitary Ensemble (GUE)~\cite{TW,leshouches}.

The problem of finding the longest common subsequence in a pair of sequences drawn from alphabet of
$c$ letters is explicitly formulated as follows. Consider two sequences $\alpha=\{ \alpha_1,
\alpha_2,\dots, \alpha_i\}$ (of length $i$) and $\beta=\{\beta_1, \beta_2, \dots,
\beta_j\}$ (of length $j$). For example, $\alpha$ and $\beta$ can be two random strings of $c=4$
base pairs A, C, G, T of a DNA molecule, e.g., $\alpha=\{\rm A, C, G, C, T, A, C\}$ with $i=6$ and
$\beta=\{\rm C, T, G, A, C\}$ with $j=5$. Any subsequence of $\alpha$ (or $\beta$) is an ordered
sublist of $\alpha$ (or $\beta$), i.e. subsequences which need not be consecutive. For example,
$\{\rm C, G, T, C\}$ is a subsequence of $\alpha$, but $\{\rm T, G, C\}$ not. A common subsequence
of two sequences $\alpha$ and $\beta$ is a subsequence of both of them. For example, the
subsequence $\{\rm C, G, A, C\}$ is a common subsequence of both $\alpha$ and $\beta$. There are
many possible common subsequences of a pair of sequences. The aim of the LCS problem is to find the
longest of them. This problem and its variants have been widely studied in biology
\cite{NW,SW,WGA,AGMML,ZM}, computer science \cite{SK,AG,WF,Gusfield}, probability theory
\cite{CS,Deken,Steele,DP,Alex,KLM,Seppa} and more recently in statistical physics
\cite{Hwa,Monvel}. A particularly important application of the LCS problem is to quantify the
closeness between two DNA sequences. In evolutionary biology, the genes responsible for building
specific proteins evolve with time and by finding the LCS of the 'same' gene in different species,
one can learn what has been conserved in time. Also, when a new DNA molecule is sequenced {\it in
vitro}, it is important to know whether it is really new or it already exists. This is achieved
quantitatively by measuring the LCS of the new molecule with another existing already in the
database.

\begin{figure}[ht]
\epsfig{file=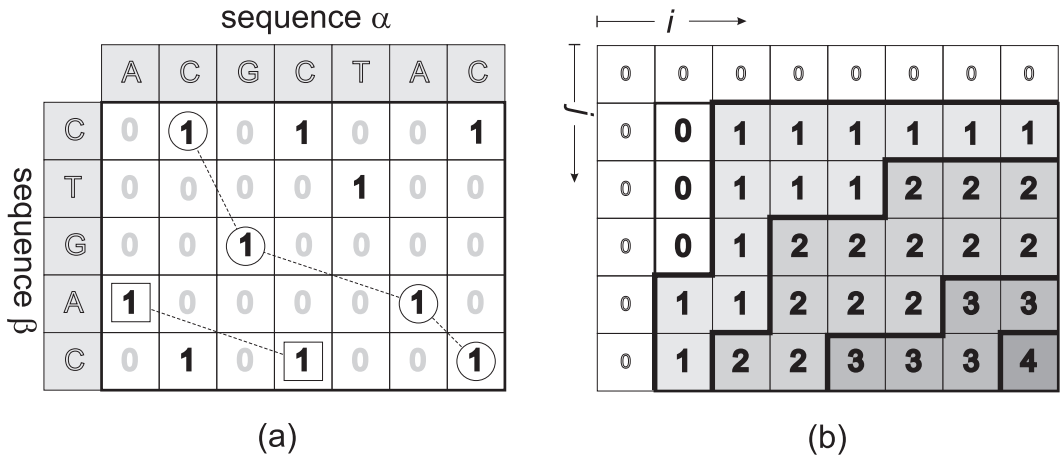,width=14cm} \caption{(a) Matrix of variables $\eta_{i,j}$; (b) Table of
lengths $L_{i,j}$ of all LCSs corresponding to (a).} \label{fig:1}
\end{figure}

Computationally, the easiest way to determine the length $L_{i,j}$ of the LCS of two arbitrary
sequences of lengths $i$ and $j$ (in polynomial time $\sim O(ij)$) with no cost of gaps can be
achieved using the simple recursive algorithm \cite{Gusfield,Monvel}
\be
L_{i,j} = \max\left[L_{i-1,j}, L_{i,j-1}, L_{i-1,j-1} + \eta_{i,j}\right],
\label{2:1}
\ee
subject to the initial conditions $L_{i,0}=L_{0,j}=L_{0,0}=0$, where the variable $\eta_{i,j}$ is:
\be
\eta_{i,j} = \begin{cases} 1 & \text{if characters at the positions $i$ (in $\alpha$) and $j$ (in
$\beta$) match each other}, \\ 0 & \text{otherwise} \end{cases}
\label{2:2}
\ee
In \fig{fig:1}a the matrix of the variables $\eta_{i,j}$ is shown for a particular pair of
sequences $\alpha=\{\rm A, C, G, C, T, A, C\}$ and $\beta=\{\rm C, T, G, A, C\}$ discussed above.
Any common subsequence can be represented by a {\it directed} path connecting sequentially
\{$\eta_{i_1,j_1}, \eta_{i_2,j_2}, ..., \eta_{i_s,j_s}, ..., \eta_{i_n,j_n}$\}, where
$\eta_{i_1,j_1} =\eta_{i_2,j_2} =...=\eta_{i_s,j_s}= ... = \eta_{i_n,j_n}=1$ and $i_{s+1}>i_s$,
$j_{s+1}>j_s$ for all $1\le s\le n$. Two particular realizations of common subsequences, namely
$\{\rm C, G, A, C\}$, and $\{\rm A, C\}$ are shown in \fig{fig:1}a by two broken lines connecting
'1' (the first one is the one of LCSs). In \fig{fig:1}b we show the table of all $L_{i,j}$ ($0\le
i\le 6$, $0\le j\le 5$) corresponding to the matrix $\eta_{i,j}$ (the first line and the first
column are the boundary conditions $L_{0,j}= L_{j,0}=0$). Let us note the terrace--like structure
of \fig{fig:1}b: the numbers from 0 to 4 could be viewed as different 'heights' of the terraces. We
shall address later to this representation.

For a pair of fixed sequences of lengths $i$ and $j$ respectively, the length $L_{i,j}$ of their
LCS is just a number. However, in the statistical version of the LCS problem one compares two
random sequences drawn from the alphabet of $c$ letters and hence the length $L_{i,j}$ is a random
variable. The statistics of $L_{i,j}$ has been intensively studied during the past three decades
\cite{CS,Deken,Steele,DP,Alex}. For equally long sequences ($i=j=n$), it has been proved that
$\langle L_{n,n}\rangle \approx \gamma_c n$ for $n\gg 1$, where the averaging is performed over all
uniformly distributed random sequences. The constant $\gamma_c$ is known as the Chv\'atal--Sankoff
constant which, to date, remains undetermined though there exists several bounds
\cite{Deken,DP,Alex}, a conjecture due to Steele \cite{Steele} that $\gamma_c=2/(1+\sqrt{c})$ and a
recent proof \cite{KLM} that $\gamma_c\to 2/\sqrt{c}$ as $c\to \infty$. Unfortunately, no exact
results are available for the finite size corrections to the leading behavior of the average
$\langle L_{n,n}\rangle$, for the variance, and also for the full probability distribution of
$L_{n,n}$. Thus, despite tremendous analytical and numerical efforts, exact solution of the random
LCS problem is far from being completely resolved. One feature that makes this problem particularly
complicated is that  the variables $\eta_{i,j}$ that are defined in ~(\ref{2:2}) are not mutually
independent but are  correlated. To see that consider the simple example -- matching of two strings
$\alpha={\rm AB}$ and $\beta={\rm AA}$. One has by definition: $\eta_{1,1}=\eta_{1,2}=1$ and
$\eta_{2,1}=0$. The knowledge of these three variables is sufficient to predict that the last two
letters do not match each other, i.e., $\eta_{2,2}=0$. Thus, $\eta_{2,2}$ can not take its value
independently of $\eta_{1,1},\,\eta_{1,2},\,\eta_{2,1}$. Note however that for two random sequences
drawn from the alphabet of $c$ letters, the correlations between the $\eta_{i,j}$ variables vanish
in the $c\to \infty$ limit.

A first natural question is whether the problem is solvable in the absence of correlations between
the $\eta_{i,j}$'s? This question leads to the Bernoulli Matching (BM) model which is a simpler
variant of the original LCS problem where one ignores the correlations between $\eta_{i,j}$'s for
all $c$ \cite{Monvel}. The length $L_{i,j}^{BM}$ of the BM model satisfies the same recursion
relation as in Eq.\eq{2:1} except that $\eta_{i,j}$'s are now independent and each $\eta_{i,j}$ is
drawn from the bimodal distribution:
\be
p(\eta)= \frac{1}{c}\delta_{\eta,1}+ \left(1-\frac{1}{c}\right)\delta_{\eta,0}=
\begin{cases} \frac{1}{c} & \text{for $\eta=1$}, \\ 1- \frac{1}{c} &
\text{for $\eta=0$} \, .
\end{cases}
\label{2:3}
\ee
This approximation is expected to be exact only in the $c\to \infty$ limit. Nevertheless, for
finite $c$, the results on the BM model can serve as a useful benchmark for the original LCS model
to decide if indeed the correlations between $\eta_{i,j}$'s are important or not. Progress has been
made for the BM model which, though simpler than the original LCS model, is still nontrivial. The
average matching length $\langle L_{n,n}^{BM} \rangle$ in the BM model, for large sequence lengths,
$n$, was first computed by Seppalainen~\cite{Seppa} using probabilistic method and it was shown
that $\langle L_{n,n}^{BM} \rangle\approx \gamma_c^{BM} n$ for $n\gg 1$ where $\gamma_c^{BM}=
2/(1+\sqrt{c})$, same as the Steele's conjectured value, $\gamma_c$, for the original LCS model.
Later the same result was rederived~\cite{Monvel} in the physics literature using the cavity method
of the spin glass physics. Recently, in ~\cite{MN2}, two of us derived the asymptotic limit law for
the distribution of the random variable $L_{n,n}^{BM}$ and showed that for large $n$
\begin{equation}
L_{n,n}^{BM}\to \gamma_c^{BM} n + f(c)\, n^{1/3}\, \chi   \, , \label{asymp1}
\end{equation}
where $\gamma_c^{BM} = 2/(1+\sqrt{c})$ and $\chi$ is a random variable with a $n$--independent
distribution, ${\rm Prob} (\chi\le x)= F_{\rm TW}(x)$ which is the well studied Tracy--Widom
distribution for the largest eigenvalue of a random matrix with entries drawn from a Gaussian
unitary ensemble \cite{TW}. For a detailed form of the function $F_{\rm TW}(x)$, see \cite{TW}. We
also have shown that for all $c$,
\begin{equation}
f(c)=\frac{c^{1/6}(\sqrt{c}-1)^{1/3}}{\sqrt{c}+1} \, .
\label{fc1}
\end{equation}
This allowed us to calculate in \cite{MN2} the average length $L_{n,n}$ including the subleading
finite size correction term, as well as the variance of $L_{n,n}^{BM}$ for large $n$,
\begin{eqnarray}
\langle L_{n,n}^{BM}\rangle &\approx & \gamma_c^{BM} n + \left<\chi\right> f(c)
n^{1/3} \nonumber \\
{\rm Var}(L_{n,n}^{BM}) &\approx & \left(\langle\chi^2\rangle-{\langle\chi\rangle}^2\right)\,
f^2(c)\, n^{2/3} \, ,
\label{eq:expvar}
\end{eqnarray}
where we have used the known exact values \cite{TW}, $\langle \chi\rangle= -1.7711\dots$ and
$\langle \chi^2\rangle- {\langle \chi\rangle}^2= 0.8132\dots$. The recursion relation \eq{2:1} can
also be viewed as a $(1+1)$--dimensional directed polymer problem \cite{Hwa,Monvel} and some
asymptotic results (such as the $O(n^{2/3})$ behavior of the variance of $L_{n,n}$ for large $n$)
can be obtained using the arguments of universality \cite{Hwa}. However the limiting
Tracy-Widom distribution and the associated exact scale factors in Eqs. (4-6) derived
in \cite{MN2} can not be obtained simply from the universality arguments.

\begin{figure}[ht]
\epsfig{file=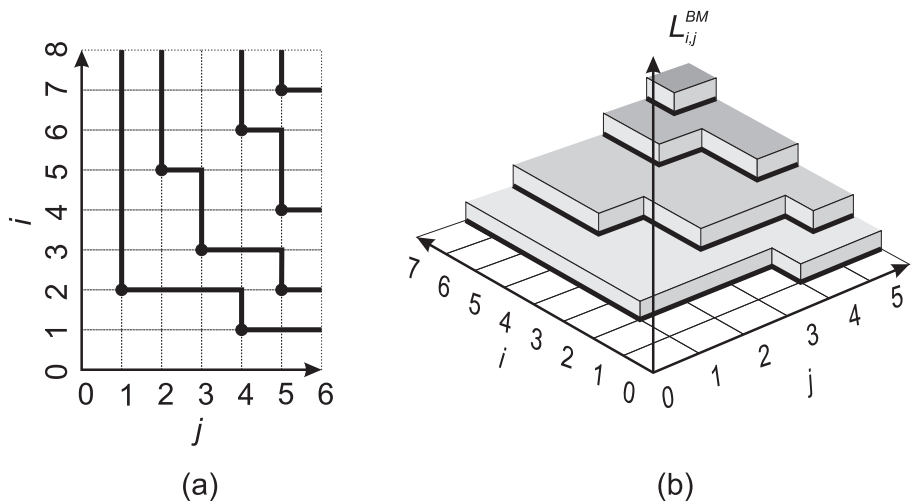,width=12cm} \caption{Bernoulli Matching model: (a)  Level lines. This is
just \fig{fig:1}b rotated by $90^o$. Note that adjacent level lines have no common segments or
edges; (b) $3$-D 'terrace--like' representation of the heights.}
\label{fig:2}
\end{figure}

As it has been mentioned above, the level structure depicted in \fig{fig:1}b can be viewed as
3--dimensional terraces. Namely, let us add one extra 'height'  dimension to the system of level
lines in \fig{fig:1}b. Each time when we cross the level line constructed via the recursion
algorithm \eq{2:1}, we increase the height by one. Hence, the length $L_{i,j}^{BM}$ can be
interpreted as the height of a surface above the 2D $(i,j)$ plane. Considering the 2D projection of
the level lines separating the adjacent terraces in \fig{fig:1}b, we can note that the rule
\eq{2:1} prohibits the overlap of these level lines, i.e., different level lines cannot have common
segments or edges --- see \fig{fig:2}a,b. The resulting 3--dimensional system of  terraces  shown
in \fig{fig:2}b is in one-to-one correspondence with the 2--dimensional system of level lines in
\fig{fig:2}a. Note that a similar (but not identical) model with terrace structures and the
associated level lines also appeared in an anisotropic 3D directed percolation model~\cite{RD},
which in turn is also related to the directed polymer problem studied by
Johansson~\cite{Johansson}. However, the levels lines in the directed percolation model can
overlap. In contrast, the level lines in our model do not overlap. These two models are
nevertheless related by a nonlinear transformation~\cite{RD,MN2}. The connections among all
mentioned and some other related models is briefly discussed in the conclusion.

All previously derived results \eq{asymp1}--\eq{eq:expvar} deal with the height statistics in the
3--dimensional terrace representation of Bernoulli Matching model. In terms of the level lines in
the projected 2D plane (see \fig{fig:2}a), the height $L_{i,j}^{BM}$ is just the total number of
level lines within the rectangle of side lengths $i$ and $j$. Note that any configuration of the
level lines in the 2D plane has an associated statistical weight (see later for details). These
weights are such that one can interpret a projected 2D configuration of the BM model as a 5--vertex
model. This 5--vertex model is an interesting model in its own right (apart from its connection to
the BM model) and it is natural  to study the statistical properties of various objects associated
with this two dimensional 5--vertex model. One such quantity is the total number of level lines
inside the rectangle with sides $i$ and $j$ that translates into the height in the BM model.
Similarly, there are other random variables such as the total number of corners and the total
number of horizontal segments in the rectangle of sides $(i,j)$ that have not been studied before,
and that have nontrivial and interesting statistical properties. For example, the 'left' corners
shown by the big dotted points in \fig{fig:3}a are the so called 'nucleation' centers for the
terraces and play an important role in the mapping between the BM model and the so called 'Longest
Increasing Subsequence' (LIS) problem \cite{MN2}. In the limit of large number of letters, $c$
($c\to \infty$), one can show that these nucleation centers are Poisson distributed in the 2D plane
with a uniform density $\rho_c=1/c$~\cite{MN2}. However, for finite $c$, the statistics of the
number of nucleation centers is, to our knowledge, still unknown. In particular, one would expect
that even the average density $\rho_c(x,y)$ of the nucleation centers for finite $c$ is nonuniform
in the $(x,y)$ plane and has a nontrivial form that reduces to the uniform value $\rho_c=1/c$ in
the $c\to \infty$ limit. In  the present paper, we shall  calculate explicitly the  average density
$\rho_c(x,y)$ of the nucleation centers for finite $c$ using the Bethe ansatz technique.

The outline of this paper is as follows. In section~\ref{section:mapping}, we explain the precise
mapping between the Bernoulli Matching model and the five vertex model and we  recall the
associated Bethe equations. We then  solve the  Bethe equations in cylindrical geometry. This
allows us in section~\ref{sect:5:2:1}  to calculate the mean flux of the word lines in the
5--vertex model as well as the associated large deviation function. Reverting  to the  Bernoulli
Matching model, our calculation leads to an independent derivation of the expectation of the
Longuest Common Subsequence. In section~\ref{sect:5:2:2}, we determine the full statistics of the
number of 'left' corners. In terms of the terrace model these  left corners play the role of
nucleation centers: our approach allows us to calculate the mean number of such centers. Concluding
remarks and a flowchart of various models related to the  Bernoulli Matching model are given in
section~\ref{sect:Conclusion}.

\section{Bernoulli Matching as a  5--vertex model:  Bethe equations}
\label{section:mapping}

The statistical weight of a projected 2D configuration of lines of Bernoulli Matching model
depicted in \fig{fig:2} is the product of weights associated with the vertices on the 2D plane. Let
$W(C)$ be the total weight of a full 2D configuration of the system of lines in \fig{fig:2}a.
Note that a new terrace is nucleated with probability $p=1/c$
and hence we associate a weight $p$ with each nucleation center. These nucleation
centers are the `left' corners of the lines shown by the big dotted points in
\fig{fig:2}a. On the other hand, each empty vertex in \fig{fig:2}a has a weight
$q=1-p$ (corresponding to a mismatch in the BM model). The other filled vertices that
are not `left' corners have an associated weight of $1$ each. Thus, we may write
the total weight of a configuration $C$ of vertices in \fig{fig:2}a as
\be
W(C)=p^{N_c}\,q^{N_e} \,\,\, \hbox{ with } \,\,\,\, p = \frac{1}{c} \, ,
\label{eq:w}
\ee
where $N_c$ and $N_e$ are the numbers of `left' corners (shown by big dotted
points in \fig{fig:2}a) and empty vertices respectively.

\begin{figure}[ht]
\centerline{\epsfig{file=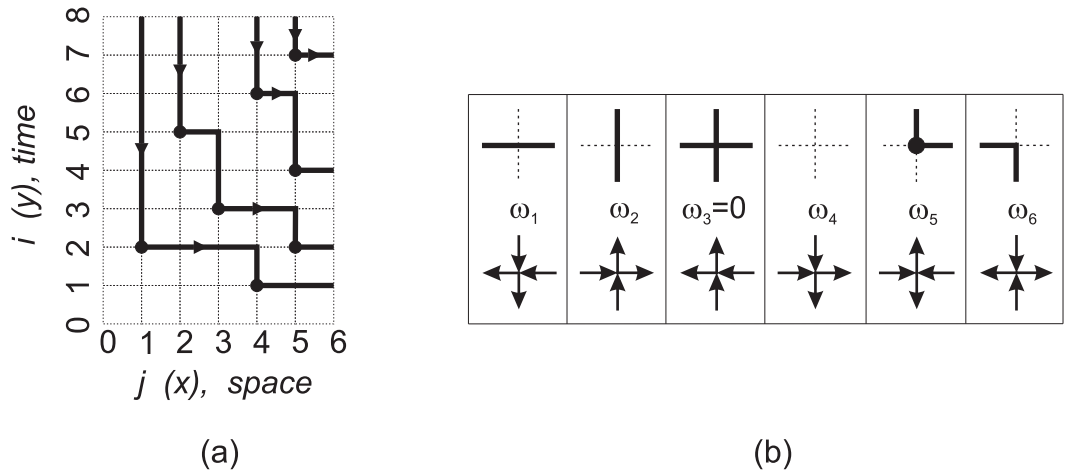, width=14cm}} \caption{(a) System of 2D level lines separating
adjacent terraces and the left corners depicted by big dots, (b) the weights of the associated
vertices. Note that the labels on the axes of Fig. (3a) correspond to the labels in \fig{fig:2}b,
but the 'time' actually runs vertically downwards. The labels (the numbers) on the vertical axis do
not correspond to time.}
\label{fig:3}
\end{figure}

One can view the configuration of trajectories in Fig.\ref{fig:2}a as an assembly of world lines of
hard--core particles moving on a one--dimensional lattice. In this alternative representation the
horizontal ($x$) and the vertical ($y$) axes in \fig{fig:3}a are correspondingly the 'time' and the
'space' directions of the 1D lattice gas picture. The dynamics of these hard--core particles is as
follows. Particles are moving along the lines as indicated in \fig{fig:3} and each particle tries
to jump to any of the empty slots available to it before the location of the next particle to its
right. Let $P(s|m)$ denote the probability that a particle hops $s$ steps, given that the next
particle to its right is at a distance $m$. Thus there are $(m-1)$ holes between the two particles
and therefore $s=0,1,2,\dots, (m-1)$. The eq.\eq{eq:w} for the total weight $W(C)$ of the
configuration dictates the following choice of this hopping probability:
\be
P(s|m)= p^{1-\delta_{s,0}}q^{m-1-s} \, .
\label{eq:hp}
\ee
One can easily check that the probability $P(s|m)$ is properly normalized: $P(k|m)=\sum_{k=0}^{m-1}
P(k|m)=1$.

There are five types of possible vertices with nonzero weights as shown in \fig{fig:3}b. Since the
level lines never cross each other, the weight $\omega_3$ is always zero. As it is seen from
definition of vertices in \fig{fig:3}b, we draw a solid line for the arrow $ \rightarrow $ or
$ \downarrow $. Otherwise we leave a bond empty. These rules slightly differ from the standard
notations in Baxter's book \cite{baxter}, but correspond to the notations used in the paper
\cite{noh}
where the Bethe ansatz equation has been considered for the 5--vertex model. Suppose now that the
weights $\{\omega_1,...,\omega_6\}$ are as follows:
\be
\omega_1=1\,; \quad \omega_2=1\,; \quad  \omega_3=0\,; \quad  \omega_4=q=1-p\,; \quad
\omega_5=p\,;
\quad \omega_6=1  \, .  \label{5:1}
\ee
Let us note that in all Bethe equations below the corner weights $\omega_5$ and $\omega_6$ always
enter in the combination $\omega_5 \omega_6$. Hence only the product weight $\omega_5 \omega_6=p$
is properly defined. Henceforth we shall call the 'left' corners (shown by the big dotted points in Fig.
(3a)) as the 'corners' on which we will mainly focus. We are not interested here in the 'right'
corners that correspond to the termination of the horizontal part of each world
line.

We first consider our vertex model in a cylindrical geometry with $N$ sites in each row and we
assume there are infinite number of rows. As usual, the number $m$ of 'up arrows'  $(\uparrow)$ in
each row is conserved. Because of the one-to-one correspondence of 'up arrows'  and 'solid' vertical
bonds in a row, $m$ is the number of solid vertical bonds and $N-m$ is the number of empty vertical
bonds in a row.

The grand canonical partition function, $Z(p)$, of a 5--vertex model reads
\be
Z(p)=\sum_{\rm conf} \omega_1^{N_h} \omega_2^{N_v}\omega_4^{N_e} (\omega_5 \omega_6)
^{N_c} \, .  \label{5:2}
\ee
where $N_h$, $N_v$, $N_e$, $N_c$ are correspondingly the numbers of: horizontal and vertical bonds,
empty vertices and corners (by corners we mean only `left' corners shown by the big dots in Fig.
(3a)) for any particular configuration of the lines (or equivalently of the arrows). The summation
in \eq{5:2} runs over all available configurations of these world lines, satisfying the
non-crossing and non-overlapping constraints.

The standard Bethe equation for the roots $z_j$ of the 5--vertex model is as follows
 (see \cite{noh} for details):
\be
z_j^N=(-1)^{N-m-1} \prod_{i=1}^{N-m} \frac{1-\Delta z_j}{1-\Delta z_i}\,; \qquad
(j=1,2,...,N-m) \label{5:3}
\ee
where
\be
\Delta=\frac{\omega_1 \omega_2-\omega_5\omega_6}{\omega_2\omega_4} \, .  \label{5:4}
\ee
The highest eigenvalue, $\Lambda_m$, can be written in two equivalent  forms:
\be
\Lambda_m=\omega_1^{N-m} \prod_{^j=1}^{m} \frac{\omega_5\omega_6}{\omega_1
z_j-\omega_4} + \omega_4^{N-m} \prod_{j=1}^{m}\frac{\omega_2\omega_1 z_j -
\omega_2\omega_4-\omega_5\omega_6 z_j}{\omega_1 z_j-\omega_4} =
 \omega_2^m \omega_4^{N-m} \prod_{j=1}^{N-m} \left(1+\frac{\omega_5 \omega_6}{\omega_2
\omega_4}z_j \right)  \, .
 \label{6:1}
\ee
 The  second expression  will be   more convenient for our further computations.
Given  the weights \eq{5:1},   we find
\be
\Delta =\frac{1-p}{q}=1  \, . \label{5:5}
\ee
Hence we obtain  the following Bethe equations
\be
z_j^N=(-1)^{N-m-1} \prod_{i=1}^{N-m} \frac{1-z_j}{1-z_i}  \, . \label{5:6}
\ee
And $\Lambda_m$ is given by Eq.\eq{6:1}:
\be
\Lambda_m=(1-q)^m \prod_{j=1}^m \frac{1}{z_j-q} + q^N \prod_{j=1}^m \frac{z_j-1}{z_j-q} =
\prod_{j=1}^{N-m}(1+p z_j)  \, .
\label{5:8}
\ee
with $z_j$ $(j=1,...,N-m)$  defined by \eq{5:6}.

Equations \eq{5:6} and \eq{5:8} are almost identical to the Bethe equation for roots and to the
expression for the highest eigenvalue of the transfer matrix of the  totally asymmetric exclusion
process (TASEP) \cite{mallick}. In the context of the exclusion process, these  equations have been
studied by many  authors \cite{derrida,dhar,gwaspohn,derrida,mallick2} (for a recent review see
\cite{mallick3}).

\section{Analysis  of the Bethe equations}
\label{sect:5:2:1}

\subsection{Statistics of the world lines: the flux}

The averaged 'flux', $\bar{\Phi}$, in the system of world lines shown in \fig{fig:3}a is equal to
the typical length (normalized per $N$) of a horizontal segment, $\la N_h \ra$, between left and
right corners averaged over all configurations. Hence, to study  the statistics of  ${\Phi}$, we
add an extra weight $e^{\mu}$ to each horizontal bond and each essential corner.  We thus define
the partition function $Z(p,\mu)$ with the following collection of weights (compare to \eq{5:1}):
\be
\omega_1=e^{\mu}\,; \quad \omega_2=1\,; \quad  \omega_3=0\,; \quad \omega_4=q=1-p\,;
\quad \omega_5 \omega_6=p e^{\mu}  \, . \label{6:2}
\ee
The mean value $\la N_h \ra$ can  then be computed in a standard way
\be
\bar{\Phi}\equiv \frac{\la N_h \ra}{N} = \frac{\disp \sum_{\rm conf} N_h\; (1-p)^{N_e}
(e^{\mu})^{N_h} (p e^{\mu})^{N_c}} {\disp N \sum_{\rm conf}  (1-p)^{N_e} (e^{\mu})^{N_h} (p
e^{\mu})^{N_c}} = \frac{1}{N}\frac{\partial}{\partial \mu} \ln Z(p,\mu)\bigg|_{\mu=0}  \, .
 \label{6:3}
\ee
The fluctuations of the average flux can be derived in the similar way:
\be
{\rm Var} (\Phi) \equiv \frac{\la N_h^2 \ra}{N^2} - \frac{\la N_h \ra^2}{N^2} = \frac{1}{N^2}
\frac{\partial^2}{\partial \mu^2} \ln Z(p,\mu) \bigg|_{\mu=0}  \, .  \label{6:4}
\ee

The particular choice of weights \eq{6:2} leads to the following expression for
$\Delta$ defined  in \eq{5:4}
\be
\Delta=\frac{\omega_1 \omega_2-\omega_5\omega_6}{\omega_2\omega_4} = e^{\mu}  \, ,
\label{6:5}
\ee
 while the general form of Bethe equation \eq{5:3} remains
unchanged. The highest eigenvalue $\Lambda_n$ reads now
\be
\Lambda_m=\omega_2^m \omega_4^{N-m} \prod_{j=1}^{N-m} \left(1+\frac{\omega_5
\omega_6}{\omega_2 \omega_4}z_j \right)=\prod_{j=1}^{N-m} \left(1-p+p e^{\mu} z_j
\right)  \, .  \label{6:6}
\ee

We proceed further using the technique of analyzing Bethe equations proposed in
\cite{derrida}. Making the change of variables
\be
y_j =  z_j\, \Delta - 1 = z_j\, e^{\mu} - 1   \, ,   \label{6:7}
\ee
we can seek the solution of Bethe equation \eq{5:3} rewritten for $y_j$:
\be
(y_j+1)^{-N} y_j^Q = \Delta^{-N} (-1)^{Q-1} \prod_{i=1}^Q y_i \label{6:8}  \, ,
\ee
in the form
\be
y_j = B^{1/Q} e^{2\pi i j/Q} (1+y_j)^{N/Q}   \, , \label{6:9}
\ee
where $B$ is a constant that will be determined self-consistently,  and $Q=N-m$. We arrive finally at the
system of equations for the parametric determination of the partition function
$Z(p,\mu) = (\Lambda_m)^N$ in the thermodynamic limit $N \to\infty$:
\be
\left\{\begin{array}{rcl} \disp \frac{1}{N} \ln Z(p,\mu) & = & \disp \sum _{j=1}^Q
\ln (1+p y_j) \medskip \\ Q \mu & = & \disp \sum_{j=1}^Q \ln (1+ y_j)
\end{array} \right.
\label{6:10}
\ee
 Equations \eq{6:9}--\eq{6:10} can be rewritten in a closed
form using the standard residue formula---see, for example \cite{complex}:
\be
\left\{\begin{array}{rcl} \disp \frac{1}{N} \ln Z(p,\mu) & = & \disp \frac{1}{2\pi
i} \oint \sum _{j=1}^Q \ln (1+p y) \left(1-\frac{N}{Q} \frac{y}{y+1}\right) \frac{d
y}{y-B^{1/Q} e^{2\pi i j/Q} (1+y)^{N/Q}} \medskip \\
Q \mu & = & \disp \frac{1}{2\pi i} \oint \sum _{j=1}^Q \ln (1+ y)
\left(1-\frac{N}{Q} \frac{y}{y+1}\right) \frac{d y}{y-B^{1/Q} e^{2\pi i j/Q}
(1+y)^{N/Q}}
\end{array} \right.
\label{6:11}
\ee

Solving \eq{6:11} we express $\ln Z(p,\mu)$ as a series expansion
\be
\left\{\begin{array}{rcl} \disp \frac{1}{N} \ln Z(p,\mu) & = & \disp p
\sum_{k=1}^{\infty} B^k \frac{{\cal F}(N k, Q k)}{k} \medskip \\
\mu & = & \disp \sum_{k=1}^{\infty} B^k \frac{(Nk-1)!}{(Q m)!((N-Q)k)!}
\end{array} \right.
\label{6:12}
\ee
where
\be
{\cal F}(Nk,Qk) = \sum_{m'=0}^{Qk-1} {Nk \choose m'} (-p)^{Qk-m'-1} \, .  \label{6:13}
\ee

For the computation of the expectation and the variance of the flux it is sufficient
to cut the series \eq{6:13} at the second term.  \bs
\beqn
\frac{1}{N} \ln Z(p,\mu) = p B {\cal
F}(N,Q) + \frac{1}{2} p B^2 {\cal F}(2N, 2Q) \medskip \label{6:14a} \\
\mu = B \frac{(N-1)!}{Q!(N-Q)!} + B^2 \frac{(2N-1)!}{(2Q)!(2(N-Q))!} \, , \label{6:14b}
\eeqn
\es where
\be
\left\{\begin{array}{rcl} {\cal F}(N,Q) & \simeq & \disp {N \choose Q-1}
\frac{1}{1+\frac{1-\rho}{\rho} p} = \frac{\rho}{p+\rho q} {N \choose Q-1}
\medskip \\ \disp {\cal F}(2N,2Q) & \simeq & \disp {2N \choose 2Q-1}
\frac{1}{1+\frac{1-\rho}{\rho} p} = \frac{\rho}{p+\rho q} {2N \choose 2Q-1}
\end{array} \right.
\label{6:15}
\ee
and we have defined the density $\rho$ (of world lines) in \eq{6:15} in the limit $N\to \infty$ as
\be
\rho=\frac{m}{N}\equiv \frac{N-Q}{N}  \, .
\label{6:16}
\ee

Now we can extract the constant $B$ from the \eq{6:14b} and substitute it into \eq{6:14a}. After some
algebra
and using Stirling formula, we arrive at the expression for the free energy (mean value and
fluctuations) of the system in the thermodynamic limit $N\to\infty$ in the ensemble with fixed
density $\rho$ of world (level) lines (i.e. with fixed total number of world lines, $m=\rho N$) and
fugacity, $\mu$, of horizontal bonds (including corners):
\be
\frac{1}{N} \ln Z_h(\rho, p, \mu) \equiv \frac{1}{N} \ln Z(p,\mu) = \frac{p (1-\rho)}{p+q\rho}\,
\mu N + \frac{\sqrt{\pi}}{4}\frac{p}{p+q\rho} \frac{(1-\rho)^{3/2}}{\rho^{1/2}}\, \mu^2 N^{3/2} \, .
\label{6:17}
\ee

Differentiating Eq.\eq{6:17} with respect to $\mu$ and putting $\mu=0$, we get the expectation of
the flux $\bar{\Phi}$ in the system:
\be
\bar{\Phi} = \frac{p(1-\rho)}{p+q\rho} \, .   \label{7:1}
\ee
This expression thus reproduces the result obtained in \cite{RD} by a different method.
According to
\eq{6:4} the variance ${\rm
Var}(\Phi)$ reads
\be
{\rm Var} (\Phi)= \frac{\sqrt{\pi}}{4}\frac{p}{p+q\rho} \frac{(1-\rho)^{3/2}}{\rho^{1/2}}\,
N^{-1/2} \, .
\label{eq:var1}
\ee

More generally, eliminating $B$ between the  two equations of~(\ref{6:12}),  allows us  to
determine the moments of $\Phi$ to any desired order.

\subsection{Expected Length of the Longest Common Subsequence in Bernoulli Matching model}

Let us return to the system of  world lines  shown in \fig{fig:3}a.
Our aim in this subsection is to derive a macroscopic coarse-grained equation of state
for the average height ${\bar L}(x,y)$ in the BM model which is
precisely the average length of the longest match.
To derive this, we return to the system of  world lines  shown in \fig{fig:3}a
then closely follow a similar line of arguments that were used in \cite{RD}.
Let us first define $\partial_x
L(x,y)\equiv \frac{\partial L(x,y)}{\partial x}$ and $\partial_y L(x,y)\equiv \frac{\partial
L(x,y)}{\partial y}$. These are random variables and we would like
to first estimate their average values.
First consider
crossing the world lines by going horizontally along the $x$ direction.
The variable $\partial_x L(x,y)$ takes the values
$$
\partial_x L(x,y)=\left\{
\begin{array}{ll}
0 & \mbox{if we do not cross a level line} \medskip \\
1 & \mbox{if we cross a level line}
\end{array} \right.
$$
Thus the average value of $\partial_x L(x,y)$ is the number of lines encountered per unit distance
along the horizontal ($x$) direction. In the 1D--lattice gas language this is just the density
$\rho$ of particles.
Hence, our first relation is:
\be
\partial_x\bar{L}(x,y) = \rho \, .
\label{7:2}
\ee

Let us now try to relate the second object $\partial_y {\bar L}(x,y)$ also to the density
$\rho$. For this, let us imagine
crossing the world lines along the vertical ($y$) axis.
It is clear that one can write,
\begin{equation}
\partial_y {\bar L}(x,y)= \partial_x {\bar L}(x,y) \frac{\partial x}{\partial y}=\rho
\frac{\partial x}{\partial y}
\label{cg0}
\end{equation}
and then replace $\frac{\partial x}{\partial y}$ by
its coarse-grained value, namely,
\begin{equation}
\frac{\partial x}{\partial y}\approx \frac{\langle N_h\rangle}{N_l}
\label{cg1}
\end{equation}
where $N_l$ is the total number of world lines or equivalently the total number
of particles in the $1$-d lattice with $N$ sites. The rhs of Eq. (\ref{cg1}) is just the mean
length of the total horizontal segment in each row per world line and
hence is also the mean horizontal distance
travelled by each particle in unit time (note that one unit of time corresponds to
traversing one row in the vertical direction) and hence it
represents the macroscopic value of $\frac{\partial x}{\partial y}$ on the lhs
of Eq. (\ref{cg1}). Now, writing $N_l= \rho N$, we immediately get from Eqs. (\ref{cg1})
and (\ref{cg0})
\begin{equation}
\partial_y {\bar L}(x,y)= \frac{\langle N_h\rangle}{N}={\bar \Phi}=
\frac{p(1-\rho)}{p+q\rho}  \, .  \label{7:3}
\end{equation}
where we have used the expression of ${\bar \Phi}$ from Eq. (\ref{7:1}).

Note that the density $\rho$ in both Eqs. (\ref{7:2}) and (\ref{7:3})
is an unknown macroscopic variable, which however can be eliminated
from the two equations. This then gives the desired
equation for the
average surface height:
\be
p\left(1-\partial_x\bar{L}(x,y) -\partial_y\bar{L}(x,y)\right) = q\, \partial_x\bar{L}(x,y)
\partial_y\bar{L}(x,y)  \, .
\label{7:4}
\ee
Solving \eq{7:4}, we find
the average profile of the surface shown in \fig{fig:2}b
\be
\bar{L}(x,y)=\frac{2\sqrt{p x y} - p(x+y)}{q}  \, . \label{7:5}
\ee
Note that this result is valid only in the regime $px<y<x/p$. This is because at $y=x/p$,
$\bar{L}(x,y)=x$
and hence $\rho=1$ already achieves its maximum value. Hence, in the regime $y>x/p$, $\bar{L}(x,y)$
sticks to its value $\bar{L}(x,y)=x$. Symmetry arguments show that for $y<px$, $\bar{L}(x,y)$
sticks to the value $\bar{L}(x,y)=y$. We note that this result was first derived
in \cite{Seppa} using probabilistic methods.
Putting $x=y=n$ in \eq{7:5} and recalling  that $p=1/c$, we arrive at the expression
\be
\bar{L}(n,n)=\frac{2}{1+\sqrt{c}} n \, ,
\label{eq:cs}
\ee
which coincides with the first term of Eq.(\ref{eq:expvar}) for the expectation $\la L_{n,n}\ra$.

\section{Statistics  of corners}
\label{sect:5:2:2}

\begin{figure}[ht]
\epsfig{file=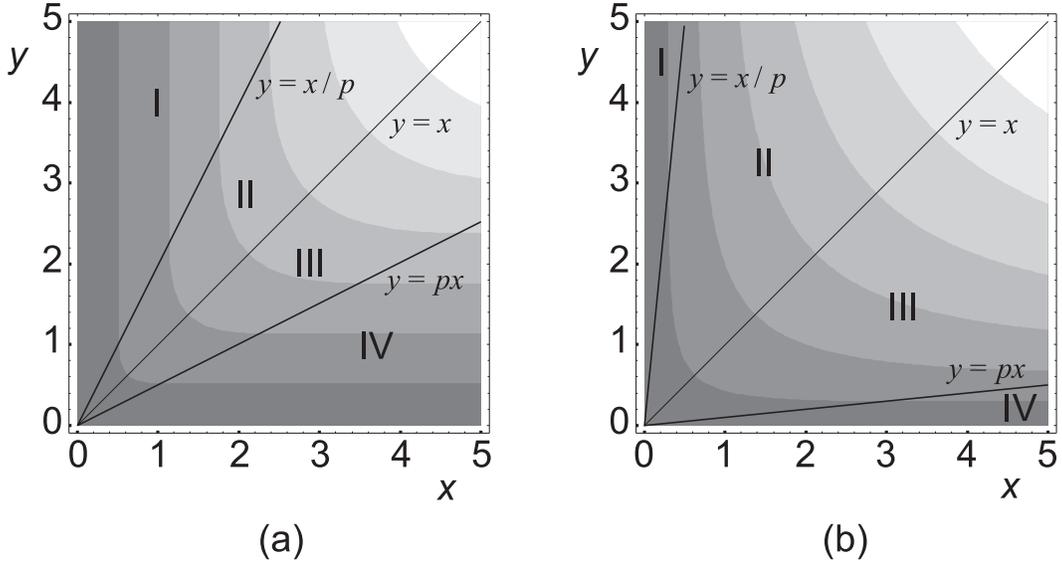,width=14cm} \caption{Averaged surface height $\bar L(x,y)$ given by
\eq{eq:height} for two different values: (a) $p=0.5$ and (b) $p=0.1$.}
\label{fig:regions}
\end{figure}

In this Section,  we compute the mean number of corners which play the role of nucleation
 centers in the terrace model.
As mentioned above,
the expression \eq{7:5} is valid in the angle $px<y<x/p$, while outside this
region the average surface height is linearly increasing along $x$ for $y\ge x/p$ and along $y$ for
$y\le px$. Hence, the complete expression for $\bar L(x,y)$ is as follows:
\be
\bar L(x,y) = \left\{\begin{array}{cl} \disp \frac{1}{q}\Big(2\sqrt{p x y} - p(x+y)\Big) & \mbox{if
$px<y<x/p$} \\ x & \mbox{if $y\ge x/p$} \medskip \\ y & \mbox{if $y\le p x$}
\end{array} \right.
\label{eq:height}
\ee
The function $\bar L(x,y)$ is depicted in \fig{fig:regions} for two different values of $p$.

Since the function $\bar L(x,y)$ is symmetric with respect to axis $y=x$, we can consider the
sector $x\le y$ only and extend the obtained results to the sector $x\ge y$ afterwards. The local
density of lines, $\rho(x,y)$, is defined in the following way
\be
\rho(x,y)=\left\{\begin{array}{ll} \disp \frac{\partial \bar L(x,y)}{\partial x}=1 & \mbox{if $x,y$
belong to the sector I, $y \ge \frac{x}{p}$} \medskip \\ \disp \frac{\partial \bar L(x,y)}{\partial
x}= \frac{\sqrt{p}}{q}\left(\sqrt{\frac{y}{x}}-\sqrt{p} \right) & \mbox{if $x,y$ belong to the
sector II, $x\le y < \frac{x}{p}$}
\end{array} \right.
\label{eq:density}
\ee

Note that in all the results that we derived in the $5$-vertex model in a cylindrical geometry we have
assumed a constant density
$\rho$ of world lines using it as a given fixed parameter of the model. Now, as seen above
in Eq. (\ref{eq:density}), for the BM model in $2$-D, the density of lines $\rho(x,y)$
is not a constant, but is a function of the space. To use the $5$-vertex results in calculating
the mean corner density in the BM model, we
will use a 'coarse-grained' description in the following sense. We consider the BM model
on a very big lattice. Now, we consider a part of these world lines over a sufficiently
big 'coarse-grained' region around the point $(x,y)$. In this `local' region, we will
consider the line density $\rho(x,y)$ to be sufficiently slowly varying function and use it as a
constant
input in the corresponding $5$-vertex model to calculate the 'local' corner density
in the region around the point $(x,y)$.

For each value of the fixed line density, $\rho(x,y)$, we need to evaluate the corner density,
$\rho_{\rm c}(x,y)$ from the 5--vertex model. The vertex weights in this case are as follows
 (compare to \eq{5:1} and \eq{6:2}):
\be
\omega_1=1; \quad \omega_2=1\,; \quad  \omega_3=0\,; \quad \omega_4=q=1-p\,; \quad
\omega_5 \omega_6=p e^{\alpha}  \, .    \label{6:18}
\ee
The mean value $\la N_c \ra$ can be computed as follows
\be
\la N_c \ra = \frac{\disp \sum_{\rm conf} N_c\; (1-p)^{N_e} (p e^{\alpha})^{N_c}}
{\disp \sum_{\rm conf}  (1-p)^{N_e} (p e^{\alpha})^{N_c}} = \frac{\partial}{\partial
\alpha} \ln Z(p,\alpha)\bigg|_{\alpha=0}  \, .    \label{6:19}
\ee
The fluctuations of the average number of corners can be derived in the similar way as the variance
of the flux. Thus, the expression for the variance ${\rm Var}(N_c)$ is as follows
\be
{\rm Var} (N_c) \equiv \la N_c^2 \ra - \la N_c \ra^2 = \frac{\partial^2}{\partial
\alpha^2} \ln Z(p,\alpha) \bigg|_{\alpha=0}  \, . \label{6:20}
\ee
The highest eigenvalue $\Lambda_m$  \eq{6:1} now reads:
\be
\Lambda_n =\prod_{j=1}^{N-m} \left(1-p+p e^{\mu} z_j
\right) = \left(\frac{q}{1-p e^{\alpha}}\right)^Q \prod_{j=1}^{Q} \left(1+p
e^{\alpha} y_j \right)   \, ,  \label{6:21}
\ee
with
\be
 y_j= z_j\, \Delta -1 = z_j\, \frac{q}{1-p e^{\alpha}} -1  \, \,\,\, \hbox{ and } \,\,\,
\Delta=\frac{\omega_1 \omega_2-\omega_5\omega_6}{\omega_2\omega_4} = \frac{1-p
e^{\alpha}}{q} \label{6:22} \, .
\ee
Solving Bethe equations, we get the  parametric system of equations defining
the free energy (compare to \eq{6:12}):
\be
\left\{\begin{array}{rcl} \disp \frac{1}{N} \ln Z(p,\alpha) & = & \disp -Q \ln
\frac{1-p e^{\alpha}}{q} + p e^{\alpha} \sum_{k=1}^{\infty} B^k \frac{{\cal F}(N k,
Q k)}{k} \medskip \\ \disp \ln \frac{1-p e^{\alpha}}{q} & = & \disp
\sum_{k=1}^{\infty} B^k \frac{(Nk-1)!}{(Q m)!((N-Q)k)!}
\end{array} \right.
\label{6:24}
\ee
with ${\cal F}(Nk, Qk)$ as in \eq{6:13}.
Proceeding  as in  Section \ref{sect:5:2:1}, we arrive at the following
expression for the free energy of the system in the ensemble with fixed density, $\rho$, and
fugacity of corners, $\alpha$:
\be
\frac{1}{N} \ln Z_{\rm c}(\rho, p, \alpha) \equiv \frac{1}{N} \ln Z(p,\alpha) = -\frac{\rho
(1-\rho)(1-p e^{\alpha})}{p e^{\alpha} +(1-p e^{\alpha})\rho}\, \eta N + \frac{\sqrt{\pi}}{2}
\frac{p e^{\alpha}}{p e^{\alpha} +(1-p e^{\alpha})\rho} \frac{(1-\rho)^{3/2}}{\rho^{1/2}}\, \eta^2
N^{3/2} \, ,
\label{6:25}
\ee
where
\be
\eta = \ln \frac{1-pe^{\alpha}}{q}      \, .   \label{6:26}
\ee
Using \eq{6:19}, \eq{6:25}--\eq{6:25} we  derive the expression for the corner density
\be
\rho_{\rm c}=\frac{\la N_{\rm c}\ra}{N^2}=\left.\frac{1}{N^2}\frac{\partial \ln Z_{\rm c}(\rho, p,
\alpha)}{\partial \alpha}\right|_{\alpha=0}    \, .
\label{eq:cden}
\ee
After simple computations, we obtain  for $\rho_{\rm c}$ the following equation
\be
\rho_{\rm c}=\frac{\rho(1-\rho)p}{p+q\rho} \, ,
\label{eq:cden2}
\ee
where for $\rho$ we should understand the local line density $\rho(x,y)$ given by
\eq{eq:density}. Substituting \eq{eq:density} into \eq{eq:cden2} we arrive at
\be
\rho_{\rm c}(x,y)=\left\{\begin{array}{ll} 0 & \mbox{for $y \ge \frac{x}{p}$} \medskip \\ \disp
\frac{p}{q^2}\frac{\left(\sqrt{\frac{y}{x}}-\sqrt{p} \right)\left(1-\sqrt{p}
\sqrt{\frac{y}{x}}\right)}{\sqrt{\frac{y}{x}}} & \mbox{for $x\le y < \frac{x}{p}$}
\end{array} \right.
\label{eq:cden3}
\ee
Since the function $\rho_c(x,y)$ is symmetric with respect to $y=x$ axis, we can straightforwardly
reconstruct the values of $\rho_c(x,y)$ in the regions III and IV from \eq{eq:cden3}.

The average number, $\la N_{\rm c}\ra$, of corners in the square box $\Omega$ of size $L\times L$
can be obtained by integrating the corner density, $\rho_{\rm c}(x,y)$ in $\Omega$ where
$\Omega$ denotes the region II and III:
\be
\la N_{\rm c}\ra = \int\limits_{\Omega} \rho_{\rm c}(x,y) dx dy = 2 \int\limits_0^L dx
\int\limits_{px}^x \rho_{\rm c}(x,y) dy = 2 \int\limits_0^L dx \int\limits_{px}^x
\frac{p}{q^2}\frac{\left(\sqrt{\frac{y}{x}}-\sqrt{p} \right)\left(1-\sqrt{p}
\sqrt{\frac{y}{x}}\right)}{\sqrt{\frac{y}{x}}} dy=\frac{L^2 p(1-\sqrt{p})^3(3+\sqrt{p})}{3q^2} \, ,
\label{eq:corners}
\ee
where $q=1-p$. It is easily seen that at $p\to 0$ the averaged number of corners, $\la N_{\rm c}
\ra$ tends to the value $p L^2$ that corresponds to the mean of the Poisson distributed corners or
nucleation centers.

We conclude  this section with the following remark. The definition of the density of level lines,
$\rho$, deserves special attention. We distinguish
the 'global', $\rho_{\rm gl}$, and 'local', $\rho_{\rm loc}$, densities, defined respectively as
follows: \bs
\beqn
\rho_{\rm gl} & = & \disp \frac{m}{N} \equiv \frac{{\bar L}(x,y)}{N}\bigg|_{x=y=N}=
\frac{2\sqrt{p}}{1+\sqrt{p}} \label{7:6a} \\ \rho_{\rm loc} & = & \disp \frac{\partial {{\bar
L}(x,y)}}{\partial x}\bigg|_{x=y=N} = \frac{\sqrt{p}}{1+\sqrt{p}}
\label{7:6b}
\eeqn
\es Note that Eqs.\eq{7:1}--\eq{7:2} are consistent with the density $\rho=\rho_{\rm loc}$ defined
in \eq{7:6b}.

\section{Conclusion}
\label{sect:Conclusion}

To summarize, in this paper we have shown how to use the Bethe ansatz technique to compute the
average number of lines and corners in a 2D 5--vertex model that originated from the Bernoulli
Matching model of the alignment of two random sequences. The asymptotic result for the average
number of lines in a square of size $(n\times n)$, which in the Bernoulli Matching model is the
average length $\langle L_{n,n}^{BM}\rangle $ of the longest match between two random sequences
each of length $n$, was already known by several alternate methods. These include a purely
probabilistic method~\cite{Seppa}, the cavity method of the spin glass physics~\cite{Monvel}, via
mapping to a lattice gas of interacting particle systems~\cite{RD} and also via a
mapping~\cite{MN2} to the Johansson's directed polymer problem~\cite{Johansson}. Here, we have
provided yet another method namely the standard Bethe ansatz technique to compute this asymptotic
result. Moreover, we were also able to compute, by this method, the average density of corners in
the $5$-vertex model which, to our knowledge, is a new result. While the number of lines has an
immediate physical meaning in the context of the sequence matching problem (namely, this is just
the number of optimal matches between two random sequences), it is difficult to relate the corner
density in the 2D plane to a directly measurable observable in the sequence matching problem. In an
indirect way the corner density is related to the degeneracy of the optimal match. However, the
corner density has a direct physical meaning in terms of the world lines of the totally asymmetric
exclusion process (TASEP). The corner density is just the total number of particle jumps per unit
time and per unit length.

\begin{figure}[ht]
\centerline{\epsfig{file=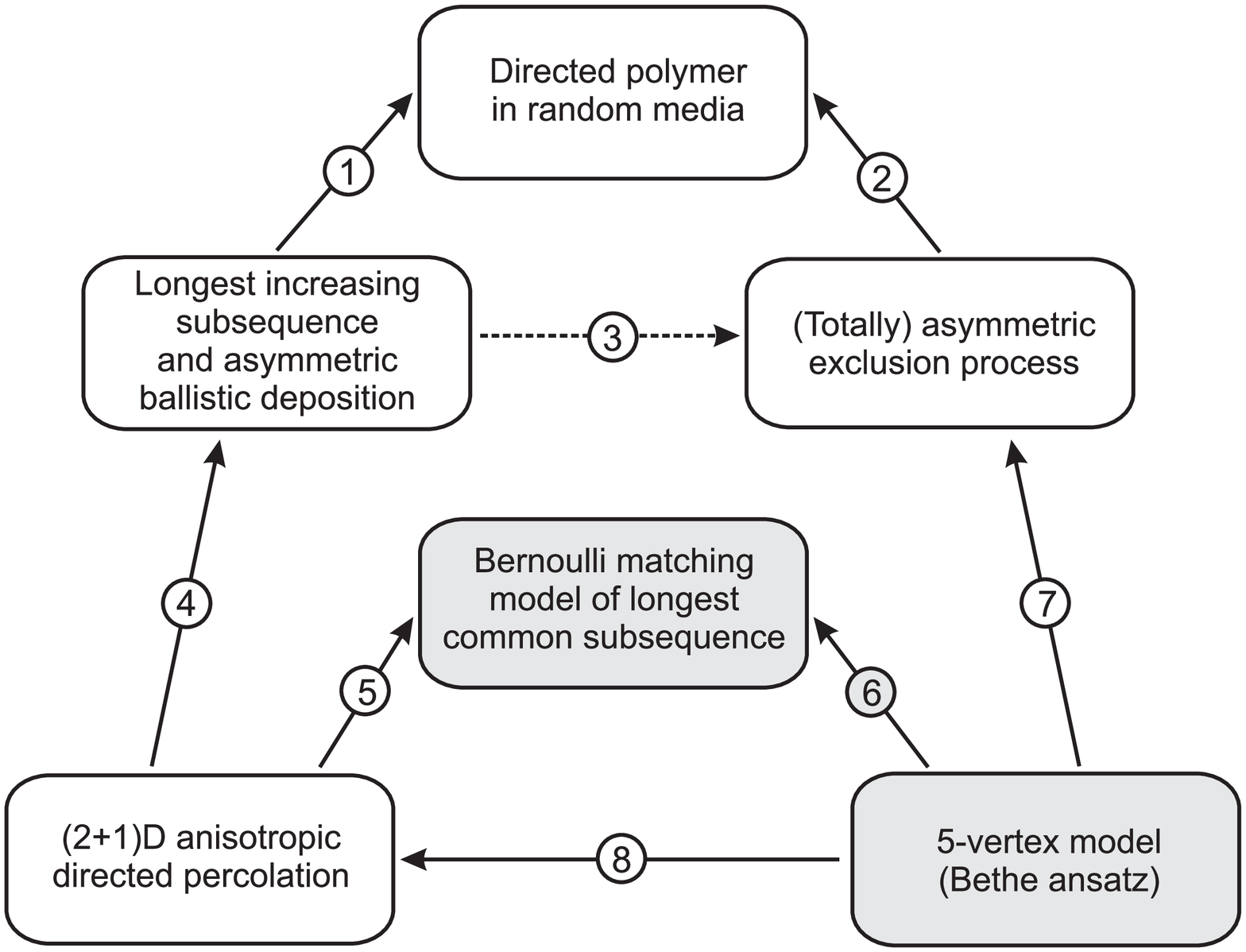, width=12cm}} \caption{A flowchart of the models related to
the Bernoulli Matching model of sequence alignment.}
\label{fig:5}
\end{figure}

There are several models and the associated techniques that are directly or indirectly related to
the Bernoulli Matching model~\cite{leshouches}. These models include the $(1+1)$-dimensional
directed polymer problem in a random medium~\cite{Johansson}, an anisotropic $(2+1)$-dimensional
directed percolation model~\cite{RD}, the longest increasing subsequence problem~\cite{LIS}, a
$(1+1)$--dimensional anisotropic ballistic deposition model~\cite{MN1} and also the 2--dimensional
5--vertex model described in this paper. They all share the common fact that there is a suitable
random variable in the model whose limiting distribution is described by the Tracy--Widom law first
appeared as the limit distribution of the largest eigenvalue of a random matrix drawn from the
Gaussian unitary ensemble. In \fig{fig:5}, we provide a flowchart of these different models and the
connections between them are depicted by numbered arrows which designate the methods of solutions
of listed problems. Below we briefly comment on these connections.

The first comprehensive solution for the distribution of the ground state energy of directed
polymer in random media (DPRM) has been obtained by Johansson in \cite{Johansson} by mapping this
model to the longest increasing (non-decreasing) subsequence (LIS) in a random sequence of
integers, known also as Ulam problem (see arrow $1$). He also discussed the possible connection
between the asymmetric exclusion process (ASEP) and LIS but since this relation has not been deeply
exploited, we have assigned to it a dashed arrow $3$. The relation between ASEP and DPRM, arrow $2$
is well-known \cite{halpinhealy}. The works \cite{RD} and \cite{MN1} have offered the possibility
for direct geometrical connection, shown by the arrow $4$ between the Tracy--Widom distribution of
LIS and the scaled height in the Anisotropic directed percolation (ADP) model. The relation between
the Bernoulli Matching (BM) model and ADP model (the arrow 5) is given by the nonlinear transform
established in \cite{MN2}. The solution of asymmetric exclusion process by Bethe ansatz, shown by
the arrow $7$ is a subject of many investigations (see \cite{mallick3} for references). As it has
been mentioned in \cite{RD}, the ADP model can be solved by mapping it to a $5$--vertex model (this
link is shown by the arrow 8) providing an alternative derivation (using Bethe ansatz) of some
results dealing with the statistics of LIS.

The shaded cells connected by the arrow 6 constitute the subject of our current work -- the use of
the Bethe ansatz technique for the 2--dimensional 5--vertex model. This thus adds one more standard
technique of statistical physics to the number of existing methods that have already been used for
investigation of problems depicted in \fig{fig:5}. A challenging forthcoming goal would be the
possibility to extract directly the Tracy--Widom distribution for the scaled height in the BM model
using the Bethe ansatz method. We note that the appearance of the Tracy-Widom law for the
distribution of total current in a fragmentation model (which includes the TASEP) was explained
before using the coordinate Bethe Ansatz method in~\cite{schutz}. This fragmentation model is to
the particle hopping model considered in our work provided one exchanges the particles and holes.
However making a link between the total current distribution in the fragmentation model and the
height distribution in the BM model is not so straightforward and remains an open question. Some
recent attempts have been made in this direction in~\cite{priez}.

Finally let us note that the BM is a special case of a more general sequence alignment problem,
where one includes a penalty (or cost) of mismatches~\cite{hwa}. Despite the fact that the
recursion relation for the cost function can still be interpreted as a generalized directed polymer
(DP) problem, the terrace-like three-dimensional structure (and hence the connection to the vertex
models) in no longer valid~\cite{hwa,tamm}. There is not clear if one could use the Bethe Ansatz
for this generalized DP problem.

\noindent{\bf Acknowledgements:} We thank D. Dhar for many useful discussions. S.M. acknowledges
the support of the Indo--French Centre for the Promotion of Advanced Research (IFCPAR/CEFIPRA)
under Project No. 3404-2. S.N. appreciates the partial support of the grant ACI-NIM-2004-243
"Nouvelles Interfaces des Math\'ematiques" (France).

\end{document}